\documentclass[aps,prc,twocolumn]{revtex4}
\usepackage{graphicx}
\usepackage{amsmath,bbm}
\usepackage{amssymb,bm}
\usepackage{array}
    
\newcommand\nd{\noindent}
\newcommand \sH {\mathcal{H}}
\newcommand \sL {\mathcal{L}}

\newcommand \sN {\mathcal{N}}
\newcommand \bx {{\bf{x} }}
\newcommand \bp {{\bf{p} }}

\newcommand \dV {{ \mathbb{V} }}
\newcommand \bigO {\mathcal{O}}
\DeclareMathOperator{\cosech}{cosech}

\begin{document}
\title{ 5-D thermal field theory, Einstein field equations and spontaneous symmetry breaking}
\author{S.~Ganesh\footnote{Corresponding author:\\Email: gans.phy@gmail.com}}
\affiliation{~~}
\begin{abstract}
	It has been shown previously, that the spatial thermal variation of a thermal medium can be recast as a variation in the Euclidean metric.
	It is now extended to temporal variations in temperature, for a non-relativistic thermal bath, which remains in local thermal equilibrium. This is achieved by examining the thermal field theory in a five-dimensional space-time-temperature. The bulk thermodynamic quantity, namely the energy density, is calculated for a neutral scalar field with a time-dependent Hamiltonian.
	Furthermore, the concept of recasting thermal variations as a variation in the metric is extended to thermal systems in a gravitational field. 
	The Einstein field equations, in the 5-D space-time-temperature, is determined. 
	It is shown that, if the scalar Lagrangian is non-minimally coupled with gravity, the resulting Ricci scalar can lead to spontaneous symmetry breaking, leading to the Higgs mechanism.
	In essence, the asymmetry in the distribution of temperature in space-time can translate to spontaneous symmetry breaking of particle fields, in a very strong gravitational field.

\vskip 0.5cm

{\nd \it Keywords } : Thermal gradient, 5-D Thermal field theory, Einstein field equation, Gravity, Spontaneous symmetry breaking, Higgs \\
{\nd \it PACS numbers } : 11.10.Wx, 04.50.Kd, 11.15.Ex

\end{abstract}

\maketitle
\section{Introduction}
\label{sec:intro}
  The thermal field theory incorporates a Euclidean space-time, obtained by an analytical continuation of time in Minkowski space-time to an imaginary time, to model thermal systems~\cite{matsubara, martin, gorkov, adas}.
Numerous research has been done by using the framework of thermal field theory in some form. Some examples include Refs.~\cite{misc1, misc2, Kraemmer, Kraemmer2, misc3, misc7, misc4, misc5, misc6, tft, epja}.
There are however thermal systems, where there are significant thermal variations in both spatial and temporal dimensions, e.g., the Quark Gluon Plasma (QGP) ~\cite{nature, naturephy}.

The idea was first mooted in Ref.~\cite{gans5}, that spatial thermal variations can be modeled by recasting the variation in the temperature as a variation in the metric. This idea was used to determine the quark anti-quark potential in a thermal system with spatial variations, using AdS-CFT correspondence. 
The thermal medium was assumed to be under local thermal equilibrium.
In Ref.~\cite{gans6}, the concept that the spatial thermal variations can be recast as a variation in the Euclidean metric, was placed on firmer grounds, by analyzing the Polyakov loop, the partition function, the correlation function, and the geodesic equation. 
It has been shown that the partition function for thermal systems with spatial thermal variations, naturally leads to the notion of a curved Euclidean space.  
While most of the analysis was carried out for a canonical ensemble, the framework was also touched upon for a grand canonical ensemble.
It is now shown, that the concept of the 5-dimensional space first introduced in Ref.~\cite{gans6}, enables the generalization of the concept to temporal variations in temperature.
The temporal variations must be sufficiently slow to maintain local thermal equilibrium.

The 5-D space-time-temperature approach further enables the extension of the concept to a thermal bath in a gravitational field. 
The Einstein field equations are determined in the 5-D space. The resultant field equations are then expressed in terms of the usual 4-D space-time covariant operators. This process of dimension reduction leads to additional terms in the Ricci tensor, and subsequently, the Ricci scalar.

There has been significant research on the source of spontaneous symmetry breaking. Reference~\cite{ssb1} has indicated that radiative corrections can be a possible source. Reference~\cite{ssb2} has explored the production of Higgs at the Large Hadron Collider (LHC).
A crucial outcome of the proposed framework is that the modified Ricci scalar may be a source of spontaneous symmetry breaking.
A negative Ricci scalar can make the term, $(m^2+R)\phi^2$, in the Lagrangian for a scalar field, become negative, i.e., $(m^2 + R)\phi^2 \rightarrow -\mu^2\phi^2$, for some real valued $\mu$.
Thus, a negative Ricci scalar curvature may be a viable candidate for the negative potential necessary for spontaneous symmetry breaking for a scalar field. The spontaneous symmetry breaking leads to the Higgs mechanism~\cite{higgs1, higgs2, goldstone}.

For the sake of clarity, let us briefly revisit some of the relevant concepts that were discussed in Ref.~\cite{gans6}. An 8-D space, $\beta^{\mu} \times x^{\nu}$, was considered to model thermal systems. 
Under conditions that the thermal system is stationary, i.e., the four-velocity of the thermal medium, $u^{\mu} = (1,0,0,0)$, it reduces to a 5-D space, $(i\beta, t, \bx)$. Moreover, if the thermal system is time-invariant, it may suffice to consider the 4-D sub-space, $(i\beta, \bx)$, leading to the imaginary time formalism. 
If the temperature varies in space, with spatial variation $s(\bx)$, i.e., $\beta \equiv \beta(\bx) = s(\bx)\beta_0$, where $\beta_0$ is a constant, then it leads to a curved Euclidean space, with the spatial variation, $h_{00} = s(\bx)^2$.
A 5-D metric tensor along with the $h_{00}$ term is shown in Eq.~\ref{eq:5Dmetric_1}.
            \begin{equation}
		    \label{eq:5Dmetric_1}
		    G^{(5)}_{ab} = 
                \left[ \begin{array}{c c}
			h_{00}& 0\\
				    0 & g^{(4)}_{\mu\nu}\\
                \end{array} \right ],
            \end{equation}
where, $g^{(4)}_{\mu\nu}$, is the usual metric tensor in 4-D space-time, resulting purely as a result of gravitational fields.
The 5-D space consideration was necessitated to allow separate metric components, $h_{00}$, due to thermal variations, and $g_{00}$, due to gravity. 
Furthermore, the 5-D space was again shown as a necessity, in order to model an external Dirac spinor, with energy $E$, traversing a thermal medium. 
The 3-momentum observable for the Dirac spinor, traversing a thermal bath, can be made manifest only with a 5-D representation of the Dirac equation, instead of the 4-D imaginary time formalism.
The conjugate momentum variables to time and temperature, namely the energy, $E$, and the Matsubara frequency, $\omega_n = \frac{C_n}{h_{00}}$, capture the intrinsic energy, $E$, of the particle, and the interaction energy with the thermal medium, $iE^c = \frac{C_n}{h_{00}}$, respectively. 
The curvature of Euclidean space was considered in the complete absence of gravity. 

In~\cite{gans6}, the analysis was primarily for a time-invariant system. However, since time and inverse temperature, $\beta$, are separate dimensions, it is tempting, to extend the formalism to time-varying systems. 
We now develop the formalism for time-varying systems.
In addition, in the current work, the curvature due to the effects of temperature variations and gravity is encapsulated in a common framework. Furthermore, the criterion for the Ricci scalar, $R$, to be negative, under the combined influence of the thermal variations and gravity, is determined.

The rest of the paper is as follows. Section~\ref{sec:timevariation}, develops the formalism for a 5-D time-varying thermal system, that remains in local thermal equilibrium.
The expectation value of energy density, for a time-varying Hamiltonian, is then determined in  Sec.~\ref{sec:evaluate}.
The 5-D Einstein field equations are developed in Sec.~\ref{sec:EFE}. The criterion for spontaneous symmetry breaking is also developed in this section.
Finally, the results are summarized in Sec.~\ref{sec:summary}.

\section{The time variation in temperature}
\label{sec:timevariation}
\begin{figure}
\includegraphics[width = 60mm,height = 40mm]{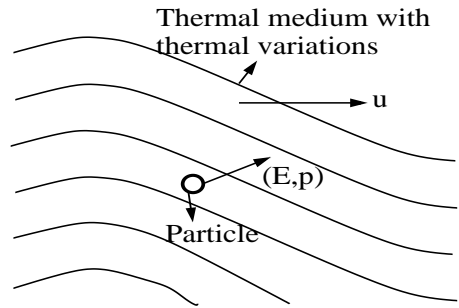}
	\caption{A thermal medium, moving with velocity $\bf{u}$, and a particle with 4-momentum, $(E,{\bf p})$, immersed in it.}
\label{fig:8dimfig}
\end{figure}

A time-varying thermal system would also have a four-velocity, $u^{\mu}$, of the thermal medium, leading to an 8-D modeling as mentioned in~\cite{gans6}. 
For reference, Fig.~\ref{fig:8dimfig} depicts a particle with 4-momentum, $(E,\bf{p})$, immersed in a thermal medium with velocity, $\bf{u}$. 
However, in the non-relativistic limit, $u^{\mu} = (1,\bf{u}) \approx (1,0,0,0)$, the analysis can be approximated by an analysis in 5-D space itself.
The 8-D space can be divided into two 4-D Lorentz invariant sub-manifolds~\cite{gans6}. Thus, a 5-D sub-space may not lend itself to a full-fledged Lorentz invariant treatment. A complete Lorentz invariant treatment would require the entire 8-D space. It is planned to extend the current framework to a complete Lorentz invariant treatment in 8-D space in the future.

In order to model partition function in a 5-D space-time-temperature, it would be required to take care that the system is not acausal or non-local in nature. This rules out an Hamiltonian of the form $H = \int \sH dtd^3x$. Instead, one may analyze the system in each time slice.
A partition function imbibes the statistical properties of the system. For a time-varying system, it is possible to take time slices 
and define the statistical properties in each time slice.
The process of taking time slices, however, does not construe that each time slice is modeled as a static system, with the effects of time derivatives being ignored. The Hamiltonian would now be time-dependent, and in general, be different from the Hamiltonian of a static system. The time-dependent Hamiltonian should capture the non-trivial effects due to the time variations.
Consequently, we define the partition function at each time slice for a time-varying system, which continues to be in local thermal equilibrium.
We now proceed on similar lines as Ref.~\cite{gans6}, for each time slice, albeit with a time-dependent Hamiltonian density.

Let us first develop the field theory for a non-interacting Lagrangian, in 5-D space-time-temperature. The Lagrangian density in a 5-D space, for a neutral scalar field would be,
\begin{equation}
\label{eq:lagrangian5D}
	\sL(\hat{\phi},\partial_a \hat{\phi}) =  \frac{1}{2} \left ( \partial_{a}\hat{\phi} \partial^{a}\hat{\phi} - m^2\hat{\phi}^2 \right ),
\end{equation}
with the index $a=0,1,2,3,4$ corresponding to the dimensions $[\tau,t,x,y,z]$. 
$\tau$ is the inverse temperature and varies from $0$ to $\beta$.
The sign convention used is (-,+,-,-,-).
This gives rise to the equation of motion:
\begin{equation}
	\partial_a\partial^a\hat{\phi} + m^2\hat{\phi} = 0
\end{equation}
The constraining 5 momentum delta function would be, $\delta( E^2 - \omega^2 - \bp^2 - m^2)$, and the 5-D integral measure is:
\begin{multline}
	\label{eq:invariant}
	\frac{1}{\beta}\sum_{n}\int \frac{d^3p}{(2\pi)^3}\frac{dE}{2\pi} 2\pi\delta(E^2 - \omega_n^2 - \bp^2 -m^2 )\\
	=  \frac{1}{\beta}\sum_{n}\int \frac{d^3p}{(2\pi)^3} \frac{1}{2E},
\end{multline}
where, the Matsubara frequency, $\omega_n = \frac{2n\pi}{\beta}$, for a boson.
Let us look at the physical interpretation of the delta function, $\delta( E^2 - \omega_n^2 - \bp^2 - m^2)$.
As mentioned in Ref.~\cite{gans6}, $E$, may be considered the original intrinsic energy of a particle, and $\omega_n = iE_c$, can be considered as the interaction energy of the particle with the thermal medium. The variable, $\omega_n$, determines the decay or enhancement of a particle wave-function with temperature (for example, the Dirac spinor in Ref.~\cite{gans6}). 
Since, $E$ and $\omega_n$ are conjugate momenta to time and inverse temperature which are orthogonal dimensions, the magnitude of the total energy is then, $\sqrt{E^2 - \omega_n^2} = \sqrt{E^2 + E_c^2}$. It is intuitive, that a particle's 3-momentum would be affected by both $E$ and $E_c$, and not just $E$. Thus, one may consider $E^2 - \omega_n^2 = \bp^2 + m^2$. A portion of the particle's original energy, $E$, is lost due to interaction with the thermal medium. This provides an intuition behind the delta function, $\delta(E^2 - \omega_n^2 - \bp^2 - m^2)$. 
A more complete physical insight, involving the delta function in 8-D space is provided in Appendix~\ref{sec:8dimsec}.

The operator for a neutral scalar field in 5-D space-time-temperature is,
\begin{multline}
	\label{eq:operator5D}
	\hat{\phi}(\bx,\tau,t) = \frac{1}{\beta}\sum_n \int \frac{d^3p}{(2\pi)^3} \frac{1}{\sqrt{2E_p}}\\
	\sum_s \left ( a^{\dagger}_{\bp,\omega_n} e^{-ipx}e^{-i\omega_n \tau} + a_{\bp,\omega_n}e^{ipx}e^{i\omega_n \tau} \right )
\end{multline}

The operator, $a^{\dagger}_{\bp,\omega_n}$, creates a particle with 3-momentum $\bp$, and Matsubara frequency $\omega_n$. 

One may premise the below commutation relation:
\begin{equation}
	\label{eq:commutation5D}
	[a_{\bp_1,\omega_{n1}}, a^{\dagger}_{\bp_2,\omega_{n2}}] = (2\pi)^3\delta^3(\bp_1 - \bp_2) \zeta(\beta)\delta_{n1,n2},
\end{equation}
where, $\zeta(\beta)$ is a scalar normalization function, and needs to be determined.
Let us define,
\begin{eqnarray}
	\label{eq:pcreation}
	\nonumber	a_{\bp}(\tau) = \sum_n f(\omega_n) a_{\bp,\omega_n}, \\
	a^{\dagger}_{\bp}(\tau) = \sum_n f^*(\omega_n) a^{\dagger}_{\bp,\omega_n}.
\end{eqnarray}
Eq.~\ref{eq:pcreation} can be interpreted in the following way. When a momentum state $|\bp\rangle$ is created, then $|\bp\rangle$ itself can be treated as a superposition of the momentum-Matsubara eigenstates $|\bp,\omega_n\rangle$, with probability amplitudes $f(\omega_n)$.
Since $f(\omega_n)$ is a probability amplitude, $\sum_n|f(\omega_n)|^2 = 1$.
Then, the equal $\tau$ commutator,
\begin{multline}
\label{eq:equaltau}
[a_{\bp_1}(\tau), a^{\dagger}_{\bp_2}(\tau)] \\
	= \sum_{n1} \sum_{n2}  [a_{\bp_1,\omega_{n1}}, a^{\dagger}_{\bp_2,\omega_{n2}}] f(\omega_{n1})f^*(\omega_{n2}) \\
	= \sum_{n1} \sum_{n2} (2\pi)^3 \delta^3(\bp_1 - \bp_2) \zeta(\beta) \delta_{n1, n2}
f(\omega_{n1})f^*(\omega_{n2}) \\
	= \sum_{n1} (2\pi)^3 \zeta(\beta) \delta^3(\bp_1 - \bp_2)|f(\omega_{n1})|^2.
\end{multline}
Since $\sum_{n1} |f(\omega_{n1})|^2 = 1$,  let us assign $\zeta(\beta) =  1$, in Eq.~\ref{eq:equaltau}, to obtain,
\begin{equation}
	\label{eq:commutator4D}
	[a_{\bp_1}(\tau), a^{\dagger}_{\bp_2}(\tau)] = (2\pi)^3\delta^3(\bp_1 - \bp_2).
\end{equation}
Thus, the usual commutation relation between the 3-momentum annihilation and creation operator is recovered.
For large $\beta$, i.e., $\beta \rightarrow \infty$, one may follow a similar procedure as above, but in the continuous domain.
The commutation relation in Eq.~\ref{eq:commutation5D}, can be seen to be,
\begin{equation}
	\label{eq:commutation5D_largeb}
	[a_{\bp_1,\omega_{n1}}, a^{\dagger}_{\bp_2,\omega_{n2}}] = (2\pi)^4\delta^3(\bp_1 - \bp_2) \delta(\omega_{n2} - \omega_{n1}),
\end{equation}
with $\zeta(\beta) = 2\pi$.
The relation between Eq.~\ref{eq:commutation5D} and Eq.~\ref{eq:commutation5D_largeb} can be understood by noting that the 4-D space-temperature manifold is a ${\mathbb R}^3\times {\mathbb S}^1$ manifold. As $\beta \rightarrow \infty$,  ${\mathbb R}^3\times {\mathbb S}^1 \rightarrow {\mathbb R}^4$.

We now proceed to determine the conjugate momenta and the Hamiltonian.
There can be a conjugate momenta w.r.t. either the time variable or the temperature variable, i.e.,
\begin{equation}
	\label{eq:pi_timetempaxis}
	\hat{\pi}_t = \frac{\delta \sL}{\delta \frac{\partial \hat{\phi}}{\partial t}};~~~
	\hat{\pi}_{\beta} = i\frac{\delta \sL}{\delta \frac{\partial \hat{\phi}}{\partial \tau}}.
\end{equation}
The corresponding Hamiltonian densities are:
\begin{equation}
	\label{eq:hamil_bothaxis}
	\sH_t = \hat{\pi}_t \frac{\partial \hat{\phi}}{\partial t} - \sL;~~
	\sH_{\beta} = -i\hat{\pi}_{\beta} \frac{\partial \hat{\phi}}{\partial \tau} - \sL.
\end{equation}
They would obey the evolution equations:
\begin{equation}
	i\frac{\partial \hat{\phi}}{\partial t} = [\hat{\phi},H_t];~~~
	\frac{\partial \hat{\phi}}{\partial \tau} = [\hat{\phi},H_{\beta}].
\end{equation}

Since we are modeling a thermal system, and are interested in evolution in $\tau$, the main object of interest would be $H_{\beta}$ and $\pi_{\beta}$. For convenience, we now drop the subscript $\beta$. In the rest of the paper, unless otherwise mentioned, $H$ and $\hat{\pi}$ refer to $H_{\beta}$ and $\hat{\pi}_{\beta}$.
We now follow a similar procedure as Ref~\cite{gans6,kapusta}, albeit modified for a 5-D space with thermal variations.

Let $\phi(\bx)$ and $|\phi(\bx)\rangle$ be the eigenfunction and the eigenket of the Schrodinger picture field operator $\hat{\phi}(\bx,0,0)$, while, $\pi(\bx)$ and $|\pi(\bx)\rangle$ be the eigenfunction and the eigenket of the conjugate momentum field operator $\hat{\pi}(\bx,0,0)$. In other words,
\begin{eqnarray}
\nonumber       \hat{\phi}(\bx,0,0)|\phi\rangle = \phi(\bx)|\phi\rangle,\\
        \hat{\pi}(\bx,0,0)|\pi\rangle = \pi(\bx)|\pi\rangle.
\end{eqnarray}
The eigenkets, $|\phi \rangle$ and $|\pi \rangle$, obey the following relation:
\begin{equation}
        \langle \phi| \pi \rangle = \exp\left (i\int d^3x \pi(\bx)\phi(\bx) \right ).
\end{equation}

For a time-dependent system, the time-dependent Hamiltonian can be written as:
\begin{equation}
	H(t) = H_0 + H_I(t),
\end{equation}
where $H_0$ is the time independent part, and $H_I(t)$ be the time dependent part.
As mentioned earlier, $H_I(t)$ should capture the non-trivial effects of time variation in a time-varying system.
$H_0$ is written in terms of the Schrodinger picture operators $\hat{\pi}(\bx,0,0)$ and  $\hat{\phi}(\bx,0,0)$.
\begin{equation}
        H_0 = \int d^3x \sH_0(\hat{\pi}(\bx,0,0), \hat{\phi}(\bx,0,0)) \equiv \int d^3x \sH_0(\bx).
\end{equation}
For simplicity of notation, the form, $\sH_0(\bx)$, is used as a representation of $\sH_0(\hat{\pi}(\bx,0,0), \hat{\phi}(\bx,0,0))$.

The free Hamiltonian density, corresponding to the Lagrangian in Eq.~\ref{eq:lagrangian5D}, would then be,
\begin{equation}
	\label{eq:Hamiltonian5D}
		\sH_0 = \frac{1}{2}\left ( \hat{\pi}^2 + (\nabla \hat{\phi})^2 - \left ( \frac{\partial \hat{\phi}}{\partial t} \right )^2 + m^2\hat{\phi}^2 \right ).
\end{equation}
But, $\frac{\partial \hat{\phi}}{\partial t} = -iE\hat{\phi}$. In a gas composed of scalar fields, which is equilibrated, $E \rightarrow 0$, as only the ensemble interaction energy, captured by the Matsubara frequency, $\omega_n$, is non zero~\cite{gans6}.
We are then left with the standard imaginary time formalism.
On similar lines, in a vacuum, as $\beta \rightarrow \infty$, $\omega_n \rightarrow 0$. Then, the only energy left is the particle energy, $E$. The Lagrangian in Eq.~\ref{eq:lagrangian5D}, then boils down to the normal 4-D space-time Quantum Field Theory (QFT).
When both $E$ and $\omega_n$ are non-zero, Eq.~\ref{eq:lagrangian5D} can be used to model particles that are not yet fully equilibrated with the thermal medium. A case in point is the external Dirac spinor traversing a thermal medium, which was modeled in Ref.~\cite{gans6}.
The external Dirac Spinor (not equilibrated with the thermal medium), will have its intrinsic energy, $E$, as well as a non-zero $\omega_n$, due to interaction with the thermal medium.
With, $E\rightarrow 0$, the Hamiltonian density in Eq.~\ref{eq:Hamiltonian5D} becomes,
\begin{equation}
	\label{eq:hamiltonian4D}
	\sH_0 = \frac{1}{2}\left ( \hat{\pi}^2 + (\nabla\hat{\phi})^2  + m^2\hat{\phi}^2 \right ).
\end{equation}
Equations \ref{eq:commutator4D} and \ref{eq:hamiltonian4D}, 
indicate that the generalizations to 5-D, characterized by Eqs. \ref{eq:operator5D}, \ref{eq:commutation5D}, \ref{eq:pi_timetempaxis}, \ref{eq:hamil_bothaxis},
are backward compatible with the existing 4-D imaginary time formalism.
A fairly generic time-dependent Hamiltonian density can be written as:
\begin{equation}
	H_I(t) = \int \sH_I(\bx,t)d^3x,
\end{equation}
where,
\begin{equation}
	\label{eq:timedephamiltoniandensity}
	\sH_I(\bx,t) = \sum_i c_i(\bx,t)f_i(\hat{\phi},\partial_{\mu}\hat{\phi}, \hat{\pi},\partial_{\mu}\hat{\pi}),
\end{equation}
and $c_i(\bx,t)$ are arbitrary scalar functions.
However, variations in $c_i(\bx,t)$, should not be sharp enough to throw the system out of local thermal equilibrium.
In this paper, the thermodynamic properties are evaluated for the below specific cases:
\begin{enumerate}
	\item $\sH_I(\bx,t) = V(\bx,t) \hat{\phi}^2$,
	\item $\sH_I(\bx,t) = \lambda(\bx,t) \hat{\phi}^4$.
\end{enumerate}
In the case of a thermal bath composed of scalar particles, $V(\bx,t) \hat{\phi}^2$ can represent the coupling of an external field, along with its derivatives, with $\hat{\phi}^2$.

The evolution operator $U_{\beta}(H_{\beta}, \beta,t)$ provides the evolution w.r.t. $\beta$, i.e.,
\begin{equation}
	U_{\beta}(H_{\beta},\beta,t) = \exp \left ( - \int \int_0^{\beta(\bx,t)} \sH_{\beta}(\bx,t) d\tau d^3x \right ).
\end{equation}
As mentioned earlier, we drop the subscript $\beta$ from $U$, $H$ and  $\sH$, and obtain,
\begin{multline}
	\label{eq:betaevolve}
	U(H,\beta,t) = \exp \left (- \int \int_0^{\beta(\bx,t)}  \sH(\bx,t) d\tau d^3x \right ),\\
	=  \exp \left (- \int \int_0^{\beta_0} s(\bx,t) \sH(\bx,t) d\tau d^3x \right ),
\end{multline}
where, $\sH(\bx,t) = \sH_0(\bx) + \sH_I(\bx,t)$.
The integrand in the exponent of Eq.~\ref{eq:betaevolve}, indicates a volume element $d\tau d^3x$ of a 4-D slice, at time $t$, within the 5-D space, with metric, $diag[-s(\bx,t)^2,1,-1,-1,-1]$, and determinant, $\sqrt{g_5} = s(\bx,t)$. Thus, it describes a curved 5-D space-time-temperature.

We can then define the partition function $Z(\beta_0,t)$, at a time slice, $t$, in 5-D space-time-temperature as:
\begin{equation}
	\label{eq:partition}
	Z(\beta_0, t) = tr \big [\langle \phi_f| U(\sH,\beta_0,t)|\phi_0 \rangle \big ],
\end{equation}
with $|\phi_0\rangle$ and $|\phi_f\rangle$ being the eigenkets at $\tau = 0$ and $\tau=\beta_0$ respectively.

The procedure to evaluate the partition function, $Z(\beta_0, t)$, is now straight forward and similar to Ref.~\cite{gans6}.
We first evaluate $U(\sH,\beta,t)$.
Let $\beta_0$ be sliced into $N$ slices, i.e., $\beta_0 = N\Delta \beta$, with $N \rightarrow \infty$ and $\Delta \beta \rightarrow 0$. This gives,
\begin{multline}
U(\sH,\beta_0,t) = \\
	\lim_{\Delta \beta \rightarrow 0}
        \exp\left (-\sum\limits_{n} \left [ \int s(\bx,t)\sH(\bx,t) d^3x \right ] \Delta \beta  \right )
\end{multline}
\begin{equation}
\label{eq:noneucaction}
        = \lim_{\Delta \beta \rightarrow 0}
          \prod\limits_{n}\exp\left (-\left [ \int s(\bx,t)\sH(\bx,t) d^3x \right ]\Delta \beta  \right ).
\end{equation}

Let $|\pi_j\rangle$ be the $\beta$ based conjugate momentum eigenstate.
Inserting a complete set of eigenstates:
\begin{equation}
\nonumber I =  \int |\phi_j\rangle \langle \phi_j| d\phi_j \times \int |\pi_{j-1}\rangle \langle \pi_{j-1}| \frac{d\pi_{j-1}}{2\pi},
\end{equation}
between each product term in Eq.~\ref{eq:noneucaction}, and
\begin{equation}
        \nonumber I =   \int |\pi_{N-1}\rangle \langle \pi_{N-1}| \frac{d\pi_{N-1}}{2\pi};~~~
I =   \int |\phi_{1}\rangle \langle \phi_{1}| d\phi_1,
\end{equation}
in the beginning and the end of the R.H.S. of Eq.~\ref{eq:noneucaction}, respectively,
we get the expression for $K(\phi_f,\phi_0,\beta_0,t) = \langle \phi_f|U(\sH,\beta_0,t)|\phi_0\rangle $:
\begin{multline}
        \label{eq:pathintinter1}
K(\phi_f,\phi_0,\beta_0,t) = \lim_{\Delta \beta \rightarrow 0}
        \int \prod\limits_{j=1}^{N-1} \langle \phi_{j+1}|\pi_j\rangle\\
\times \langle \pi_{j}| \exp\left (-\int s(\bx,t)\sH(\bx,t) \Delta \beta d^3x \right )|\phi_j\rangle\\
\times	\langle \phi_1| \phi_0 \rangle d\phi_j \frac{d\pi_j}{2\pi}.
\end{multline}
It is possible to evaluate the expression in Eq.~\ref{eq:pathintinter1}, using the relations:
\begin{itemize}
\item
\begin{equation}
\label{eq:scalarpiphi}
        \langle \phi_{j+1}|\pi_j\rangle =  \exp \left (i\int d^3x \pi_j(\bx) \phi_{j+1}(\bx) \right ),
\end{equation}
\item
\begin{multline}
\label{eq:piexpphi}
\langle \pi_j| \exp \left (-i\int s(\bx,t)\sH d^3x \Delta \beta \right )|\phi_j \rangle   \\
        = \langle \pi_j|\phi_j \rangle \exp \left ( -i\int s(\bx,t)\sH_j d^3x \Delta \beta \right ),
\end{multline}
with, $\sH_j = \sH(\pi_j,\phi_j,t) =  \sH_0(\pi_j,\phi_j) + \sH_I(\pi_j,\phi_j,t)$,
\item and
\begin{equation}
\label{eq:phiphi}
        \langle \phi_1| \phi_0 \rangle = \prod_{\bx}\delta \left (\phi_1(\bx) - \phi_0(\bx) \right ).
\end{equation}
\end{itemize}
It is to be noted that the form of $\sH_I(\bx,t)$, used in Eq.~\ref{eq:timedephamiltoniandensity}, would satisfy the relation specified in Eq.~\ref{eq:piexpphi}. This is because, the operator arguments of $\sH_I$, namely, $\hat{\phi}(\bx)$ and $\hat{\pi}(\bx)$, continue to be Schrodinger picture operators, and the temporal aspect is captured by the scalar functions, $c_i(\bx,t)$.
If a different form of $\sH_I(\bx,t)$ is used, then Eq.~\ref{eq:piexpphi} needs to be checked for validity.
Inserting Eqs.~\ref{eq:scalarpiphi},~\ref{eq:piexpphi} and~\ref{eq:phiphi} in Eq.~\ref{eq:pathintinter1}, we obtain,
\begin{multline}
\label{eq:eucaction}
        K(\phi_f,\phi_0,\beta_0,t) =
\lim_{\Delta \beta \rightarrow 0}
        \int  \prod\limits_j d\phi_j \frac{d\pi_j}{2\pi} \exp \Bigg \{ \int d^3x\\
\times \Big [ -i\pi_j(\bx)\Big \{ \phi_{j+1}(\bx) - \phi_j(\bx) \Big \} -  s(\bx,t)\sH_j  \Delta \beta \Big ] \Bigg \}
\end{multline}
\begin{multline}
        =\lim_{\Delta \beta \rightarrow 0}
          \int \left ( \prod\limits_j  d\phi_j \frac{d\pi_j}{2\pi} \right )
\exp \Bigg \{ \sum_k\Big (  \Delta \beta \int d^3x \\
        \times  s(\bx,t)  \left [ -i \pi_k(\bx)\frac{\left \{ \phi_{k+1}(\bx) - \phi_k(\bx) \right \} }{s(\bx,t) \Delta \beta } -  \sH_k  \right ] \Big ) \Bigg \}.
\end{multline}
We recollect that, $s(\bx,t) = \sqrt{h_{00}(\bx,t)}$ (Eq.~\ref{eq:5Dmetric_1}). Further, for the metric, $G^{5} = diag[-s(\bx,t)^2,1,-1,-1,-1]$, we have, $\sqrt{|G^{5}|} = s(\bx,t)$. We denote the determinant of the Euclidean metric, $|G^{5}|$, by $g_5$, and obtain in the continuum limit:
\begin{multline}
\label{eq:pathintfinal}
        K(\phi_f,\phi_0,\beta_0,t) =\\
        \int D\phi \int \frac{D\pi}{2\pi}
	\exp \Bigg \{ \int_0^{\beta_0} d\tau \int d^3x  \sqrt{g_5(\bx,t)}\\
\times  \Big [ -i\pi(\bx,\tau)\frac{1}{\sqrt{h_{00}(\bx,t)}}D_{\tau} \phi(\bx,\tau)  \\ 
	-  \sH(\pi(\bx,\tau),\phi(\bx,\tau),t)  \Big ]  \Bigg \},
\end{multline}
where, $D_{\tau}$ is the covariant derivative w.r.t. $\tau$. For a scalar $\phi(\bx,\tau)$, $D_{\tau}\phi(\bx,\tau) = \frac{\partial \phi(\bx,\tau)}{\partial \tau}$.
Finally, the partition function becomes,
\begin{multline}
\label{eq:pathintpart}
	Z(\beta_0,t) = trK(\phi_f,\phi_0,\beta_0,t) =\\
	\int_{periodic} D\phi \int \frac{D\pi}{2\pi}
	\exp \Bigg \{ \int_0^{\beta_0} d\tau \int d^3x  \sqrt{g_5(\bx,t)}\\
\times  \Big [ i\pi(\bx,\tau)\frac{1}{\sqrt{h_{00}(\bx,t)}}D_{\tau} \phi(\bx,\tau)  \\ 
	-  \sH(\pi(\bx,\tau),\phi(\bx,\tau),t)  \Big ]  \Bigg \}.
\end{multline}

\section{Evaluation of $\langle E(t) \rangle$}
\label{sec:evaluate}
In this section, the expectation value of the energy density at different instances of time, $\langle E(t) \rangle$, is evaluated for $\sH_I = V(t)\hat{\phi}^2$, and $\sH_I = \lambda(t)\hat{\phi}^4$.
\subsection{$\langle E(t) \rangle$ when $\sH_I = V(t)\hat{\phi}^2$}
Since only $\phi^2$ is involved in the partition function, this is easily evaluated as Gaussian integrals. Using the procedure outlined in~\cite{gans6}, the expression for the energy expectation value, $\langle E(t)\rangle$ is: 

\begin{multline}
\label{eq:pathintscalarR1}
	\langle E(t) \rangle =       -\frac{\partial \ln(Z)}{\partial \beta_0} =
                 \dV \int \frac{d^3p}{(2\pi)^3} \Bigg [ \frac{\omega_p}{2} \coth\left (\frac{\beta_0\omega_p}{2} \right ) \\
	+ \left ( \langle \eta(\bx,t) \rangle  + \frac{1}{2\omega_p^2}\langle s(\bx,t) V(\bx,t)\rangle \right ) \\
\times        \Big \{   \frac{\omega_p}{2} \coth\left (\frac{\beta_0\omega_p}{2} \right ) + \frac{\beta_0 \omega_p^2}{4}
         \cosech^2 \left ( \frac{\beta_0\omega_p}{2} \right )  \Big \} \Bigg ],
\end{multline}
where, $\langle s(\bx,t)V(t)\rangle = \frac{\int d^3x s(\bx,t)V(t)}{\int d^3x}$, $s(\bx,t) = 1 + \eta(\bx,t)$, $\langle \eta(\bx,t)\rangle = \frac{\int d^3x \eta(\bx,t)}{\int d^3x}$ and $\omega_p = \sqrt{\bp^2 + m^2}$.

\subsection{$\langle E(t) \rangle$ when $\sH_I = \lambda(\bx,t)\hat{\phi}^4$}
The Lagrangian density for a neutral scalar field would now read:
\begin{equation}
\label{eq:intlagscalar}
	\sL =  \frac{1}{2} \left ( \partial_{\mu}\hat{\phi} \partial^{\mu}\hat{\phi} - m^2\hat{\phi}^2 \right ) - \lambda(\bx,t) \hat{\phi}^4.
\end{equation}
The corresponding interaction Hamiltonian is, $H_I = \lambda(\bx,t) \hat{\phi}^4$. For small $\lambda(\bx,t)$, the partition function can be written as:
\begin{eqnarray}
\label{eq:intpathintscalar1}
\nonumber       Z(\beta_0,t) =             
	\int_{periodic} D\phi \int \frac{D\pi}{2\pi} \\
	\nonumber \times \left ( 1 + \sum_r\frac{1}{r!}\left [ \int d\tau d^3x \sqrt{g_5}\lambda(\bx,t) \phi^4 \right ]^r \right ) \\
\times          \exp \Bigg \{ \int_0^{\beta_0} d\tau \int d^3x  \sqrt{g_5}
\nonumber\Big [ -i\pi(\bx,\tau)\frac{1}{\sqrt{h_{00}(\bx)}}\frac{\partial \phi(\bx,\tau)}{\partial \tau}\\
\nonumber       -  \frac{1}{2} \left ( \pi^2 + (\nabla\phi(\bx,\tau))^2 + m^2\phi(\bx,\tau)^2 \right )   \Big ]  \Bigg \}.\\
\end{eqnarray}
Then, one may represent the above equation as:
\begin{equation}
	\ln Z(\beta_0,t) = \ln(Z_0(\beta_0,t)) + \ln \left [ 1 + \sum_r\frac{1}{r!}\frac{Z_I^r(\beta_0,t)}{Z_0(\beta_0,t)} \right ]. 
\end{equation}
The calculation of $\ln(Z_0(\beta_0,t))$, at time slice $t$, is identical to the calculation of $\ln(Z_0(\beta_0))$, in Ref.~\cite{gans6}.

\begin{figure}
\includegraphics[width = 60mm,height = 40mm]{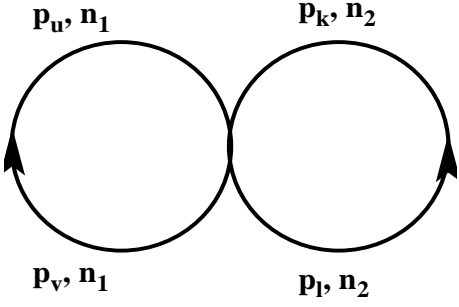}
	\caption{Feynman diagram depicting the first order $\hat{\phi}^4$ interaction.}
\label{fig:phi4int}
\end{figure}

For $r=1$, the expression for $Z^1_I$, corresponding to the Feynman diagram in Fig.~\ref{fig:phi4int} is:
\begin{eqnarray}
\label{eq:intpathintscalar3}
\nonumber       Z_I^1(\beta_0,t) =             
	\int_{periodic} D\phi \int \frac{D\pi}{2\pi} \left (\int d\tau d^3x  \sqrt{g_5}\lambda(\bx,t) \phi^4 \right ) \\
\times          \exp \Bigg \{ \int_0^{\beta_0} d\tau \int d^3x  \sqrt{g_5}
\nonumber\Big [ -i\pi(\bx,\tau)\frac{1}{\sqrt{h_{00}(\bx)}}\frac{\partial \phi(\bx,\tau)}{\partial \tau}\\
\nonumber -  \frac{1}{2} \left ( \pi^2 + (\nabla\phi(\bx,\tau))^2 + m^2\phi(\bx,\tau)^2 \right )   \Big ]  \Bigg \}.\\
\end{eqnarray}
The factor $\frac{Z_I^1}{Z_0}$, is now calculated.
The $\int D\pi$ integrals can be done in exactly the same manner as the non-interacting case. 
We consider a space of volume $\dV$, discretize it, and divide it into $M^3$ cubes of length $\Delta a$.
Further, express $\phi(\bx)$ in terms of its momentum domain representation, $\phi_{n,\bp}$, 
\begin{equation}
        \label{eq:phiexp}
\phi(\bx,\tau) = \sqrt{\frac{\beta_0}{\dV}} \sum_i\sum_{n=-\infty}^{\infty}e^{iC_n\tau} e^{i\bp_i.\bx}\phi_{n,\bp_i},
\end{equation}
where, $C_n$ is related to the Matsubara frequency, $\omega_n$, as $\omega_n = \frac{C_n}{h_{00}}$~\cite{gans6}.
We follow a procedure similar to the non-interacting case in Ref~\cite{gans6}.
Then, after replacing $\left (\frac{\Delta p}{2\pi}\right )^3$ as $\dV$ in the exponent, one may obtain $Z_I^1$ as:
\begin{multline}
\label{eq:pathintscalarZI1}
	Z_I^1 =  \Bigg [ \sN \left ( \frac{1}{2\pi}\right )^{NM^3/2} \Bigg ]\\ 
\times	\int D\phi_{n,\bp} \left ( \int d\tau d^3x s(\bx,t) \lambda(\bx,t) \phi(\bx)^4 \right )\\
\times	\exp \Bigg \{ \frac{-1}{2}
	\sum_i \sum_j \sum_{n=-\infty}^{\infty}
        \phi_{n,\bp_j}^* \Bigg [ \beta_0^2 \Big \{ (C_n^2 + \bp_i \bp_j +  m^2)\delta_{ij} \\
+ \eta_f(\bp_j - \bp_i,t)
  	\frac{1}{\dV}(-C_n^2 +\bp_i \bp_j +m^2) \Big \}
        \Bigg ]  \phi_{n,\bp_i} \Bigg \},
\end{multline}
where, we have used $\sqrt{g_5} = s(\bx,t)$, and $\sN$ is an inconsequential multiplication factor and does not affect the dynamics of the system~\cite{kapusta,Nbeta}.
The length, $\Delta a$, gets absorbed in $\sN$.
The variable, $\eta_f(\bp,t)$, is the 3-D Fourier transform of $\eta(\bx,t)$.
Based on Eq.~\ref{eq:phiexp}, $\phi(\bx)^4$ can be represented in terms of its Fourier components.
\begin{eqnarray}
	\label{eq:phi41}
	\nonumber 	\int d\tau d^3x s(\bx,t) \lambda(\bx,t)\phi^4(\bx) = \left ( \sqrt{ \frac{\beta_0}{\dV} } \right )^4 \\
\nonumber 	\times \sum_{\bp_u,n_u} \sum_{\bp_v,n_v} \sum_{\bp_k,n_k} \sum_{\bp_l,n_l} \beta_0 \delta \left (C_{n_u} + C_{n_v} + C_{n_k} + C_{n_l} \right )\\
\nonumber	\times s_{\lambda f}(-\bp_u - \bp_v - \bp_k -\bp_l,t) \phi_{n_u,\bp_u} \phi_{n_v,\bp_v}\phi_{n_k,\bp_k}\phi_{n_l, \bp_l},\\
\end{eqnarray}
where, $s_{\lambda f}(\bp,t)$ is the momentum domain representation of the product term, $s(\bx,t)\lambda(\bx,t)$.
The delta function forces, $n_u = -n_v = n_1(say)$, and $n_k = -n_l = n_2(say)$. Secondly, we make the transformation $\bp_v \rightarrow -\bp_v$ and $\bp_l \rightarrow -\bp_l$. With these, Eq.~\ref{eq:phi41}, becomes
\begin{multline}
	\label{eq:phi42}
\int d\tau d^3x s(\bx,t)\phi^4(\bx) = 3\frac{\beta_0^3}{\dV^2} \sum_{n1}\sum_{n2}\sum_u\sum_v\sum_k\sum_l\\
	\left \{ s_{\lambda,f} \left (\bp_v - \bp_u + \bp_l - \bp_k,t \right )
	\phi^*_{n1,\bp_v} \phi_{n1,\bp_u} \phi^*_{n2,\bp_l}\phi_{n2,\bp_k} \right \}.\\ 
\end{multline}

Let the exponent enclosed between $\phi_{n,\bp_j}^*$ and $\phi_{n,\bp_i}$  in Eq.~\ref{eq:pathintscalarZI1} be $M_n$, i.e., 
\begin{eqnarray}
	\label{eq:Mn}
\nonumber 	M_n = \Bigg [\beta_0^2 \Big \{ (C_n^2 + \bp_i \bp_j +  m^2)\delta_{ij} \\
+ \eta_f(\bp_j - \bp_i,t)
 `\frac{1}{\dV}(-C_n^2 +\bp_i \bp_j +m^2) \Big \}
        \Bigg ].
\end{eqnarray}
Let $\Lambda_n$ diagonalize $M_n$. Then $M_n = \Lambda_n^{-1}D_n\Lambda_n$, where $D_n$ is a diagonal matrix. Let $\psi_{n,a} = \sum_i \Lambda_{n,a,i}\phi_{n,\bp_i}$. This means  $\phi_{n,\bp_i} = \sum_a\Lambda_{n,i,a}^{-1} \psi_{n,a}$.

Inserting the above transformation of $\phi$ in  Eq.~\ref{eq:phi42}, and making use of the fact that $(\Lambda^{-1}_{n,i,a}\psi_{n,a})^* = \psi_{n,a}^*\Lambda^t_{n,i,a} = \psi_{n,a}^*\Lambda_{n,a,i}$, 
\begin{multline}
	\label{eq:phi43}
	\int d\tau d^3x\,s(\bx,t)\lambda(\bx,t)\phi^4(\bx) \\
	= 3\frac{\beta_0^3}{\dV^2} \sum_{n1}\sum_{n2}\sum_{b}\sum_a\sum_d\sum_c \psi^*_{n1,b} \psi^*_{n2,d} \\
\times \Big [ \sum_v \sum_u \sum_l \sum_k \Lambda_{n1,b,v} \Lambda_{n2,d,l}\\ 
	\times	s_{\lambda,f}(\bp_v - \bp_u + \bp_l - \bp_k,t) \Lambda^{-1}_{n1,u,a} \Lambda^{-1}_{n2,k,c} \Big ]  \psi_{n1,a}\psi_{n2,c}.\\
\end{multline}
Or,
\begin{multline}
	\label{eq:phi44}
	\int d\tau d^3x\,s(\bx,t)\lambda(\bx,t)\phi^4(\bx) \\
= 3\frac{\beta_0^3}{\dV^2} \sum_{n1}\sum_{n2}\sum_{b}\sum_a\sum_d\sum_c \\
\times	\Big ( \psi^*_{n1,b} \psi^*_{n2,d}  \Lambda^s_{bdacn_1 n_2} 
	\psi_{n1,a}\psi_{n2,c} \Big ).
\end{multline}
Equation~\ref{eq:phi44} is same as Eq.~\ref{eq:phi43}, but with a more compact notation, by introducing, $\Lambda^s_{bdacn_1 n_2}$, for the term enclosed within the $[\,]$ brackets. 

Also, we have,
\begin{equation}
	\label{eq:phi45}
	\sum_i\sum_j \phi^*_{n,\bp_j} M_{n,j,i} \phi_{n,\bp_i} = \sum_l \psi^*_{n,l}D_{n,l,l}\psi_{n,l}.
\end{equation}
Since the transformation, $\phi \rightarrow \psi$, is unitary, we can replace the measure $\int D\phi$ by $\int D\psi$. Substituting Eq,~\ref{eq:phi44}, and Eq.~\ref{eq:phi45}, in Eq.~\ref{eq:pathintscalarZI1}, we obtain:
\begin{eqnarray}
\label{eq:pathintscalarZI3}
	\nonumber                  Z_I^1 = \lambda(t) \Bigg [ \sN \left ( \frac{1}{2\pi}\right )^{NM^3/2} \Bigg ]\\ 
\nonumber \times	 \sum_{n1}\sum_{n2}\sum_{b}\sum_a\sum_d\sum_c \left (3\frac{\beta_0^3}{\dV^2} \right ) \\
\nonumber \times \int D\psi_{n,i} \left ( 
	 \psi^*_{n1,b} \psi^*_{n2,d}  \Lambda^s_{bdacn_1 n_2}    \psi_{n1,a}\psi_{n2,c} \right ) \\
\times	\exp \Bigg \{ -\frac{1}{2} \sum_i \sum_{n=-\infty}^{\infty} 
	\psi_{n,i}^*  D_{n,i,i} \psi_{n,i} \Bigg \}.
\end{eqnarray}

The functional integrals where, $b \ne a$ and $d \ne c$, are zero since the integrands are odd. The path integral then simplifies to:
\begin{multline}
\label{eq:pathintscalarZI4}
	Z_I^1 = \Bigg [ \sN \left ( \frac{1}{2\pi}\right )^{NM^3/2} \Bigg ]\\ 
\times	 \sum_{n1}\sum_{n2}\sum_a\sum_c 
	  \Lambda^s_{acacn_1 n_2}  \left ( 3 \frac{\beta_0^3}{\dV^2} \right ) \\
	\times	\int d\psi_{n1,a} \psi_{n1,a}^2 \exp \left ( -\frac{1}{2} \psi^*_{n1,a} D_{n,a,a} \psi_{n1,a} \right ) \\
	\times	\int d\psi_{n2,c} \psi_{n2,c}^2 \exp \left ( -\frac{1}{2}  \psi^*_{n2,c} D_{n,c,c} \psi_{n2,c} \right ) \\
	\times	\int_{remaining} D\psi_{n,i}
	\exp \Bigg \{ -\frac{1}{2} \sum_{\substack{(n,i) \ne\\ (n1,a),(n2,c)}} 
	\psi_{n,i}^*  D_{n,i,i} \psi_{n,i} \Bigg \}.
\end{multline}

This finally gives,
\begin{multline}
\label{eq:pathintscalarZI5}
	                  Z_I^1 = \Bigg [ \sN \left ( \frac{1}{2\pi}\right )^{NM^3/2} \Bigg ]\\ 
\times	\left ( 3\frac{\beta_0^3}{\dV^2} \right ) \sum_{n1}\sum_{n2}\sum_a \sum_c \Bigg \{ \Lambda^s_{acacn_1 n_2}\\
\times \left [ \frac{\sqrt{2\pi}}{2 D_{n1,a,a}^{3/2}} \right ]
	\left [ \frac{\sqrt{2\pi}}{2 D_{n2,c,c}^{3/2}} \right ]
	\prod_{\substack{(n,i)\ne\\ (n1,a),(n2,c)}}\left [ \frac{\sqrt{2\pi}}{2 D_{n,i,i}^{1/2}} \right ] \Bigg \},
\end{multline}
Since, $\Pi_k D_{n,k,k} = \det(D_{n}) = \det(M_n)$, it is possible to write:
\begin{eqnarray}
\label{eq:pathintscalarZI6}
	\nonumber    \frac{Z_I^1}{Z_0} = 3 \frac{\beta_0^3}{\dV^2} \sum_{n1}\sum_{n2}\sum_a \sum_c \Bigg \{\Lambda^s_{acacn_1 n_2} \\
\times 	\left [ \frac{1}{2 D_{n1,a,a}} \right ]
	\left [ \frac{1}{2 D_{n2,c,c}} \right ] \Bigg \}.
\end{eqnarray}
We now estimate,  $\Lambda^s_{acacn_1 n_2}$, to the first level of approximation. Better approximations using numerical methods or superior techniques could lead to higher accuracy.
For matrices, $M_n$, $A_n$ and $E_n$, where $M_n = A_n + E_n$, and if eigenvalues, $\lambda_A$, of $A_n$ is known, it is possible to estimate the eigenvalues, $\lambda_M$, of $M_n$ using the relation~\cite{eigen}:
\begin{equation}
	\lambda_M = \lambda_A + \frac{x^t E_n x}{x^t x} + \bigO(\|E\|^2),
\end{equation}
where, $x^t$ is eigenvector of $A_n$. Since, $\bigO(\|E\|) \sim \bigO(\|n_f\|)$, this should estimate eigenvalues of $M_n$ to order $\bigO(\|n_f\|^2)$.
The matrix, $M_n$, is given in Eq.~\ref{eq:Mn}.
Let us take $A_n$ and $E_n$ to be the matrices:
\begin{eqnarray}
	\label{eq:AE}
	A_{n,i,j} = \beta_0^2  (C_n^2 + \bp_i \bp_j +  m^2)\delta_{ij},\\
	E_{n,i,j} = \beta_0^2 \eta_f(\bp_j - \bp_i,t) \frac{1}{\dV}(-C_n^2 +\bp_i \bp_j +m^2).
\end{eqnarray}
Since $A_n$ is diagonal, $\lambda_A = \beta_0^2  (C_n^2 + \bp^2  +  m^2)$, and the eigenvectors, $x^t$, are unit vectors with only one non-zero entry.
This gives, 
\begin{multline}
	\lambda_{M;n,i}\\
	\approx \beta_0^2 \left \{ (C_n^2 + \bp_i^2  +  m^2) + \frac{\eta_f(0,t)}{\dV}(-C_n^2 +\bp_i^2 +m^2) \right \}.
\end{multline}
Thus,
\begin{multline}
	\label{eq:Daa}
	D_{n,a,a} = \lambda_{M;n,a} \\
	\approx  \beta_0^2  \left \{ (C_n^2 + \bp_a^2  +  m^2) + \frac{\eta_f(0,t)}{\dV} (-C_n^2 +\bp_a^2 +m^2) \right \}.
\end{multline}
This approximation has extracted the diagonal elements of $E$ as the contribution from $E$ to $\lambda_M$. 
Consequently, the estimated eigenvector of $M_n$ continues to be $x$, at this level of approximation.
Ergo, $\Lambda_{n,a} = \Lambda_{n,c} \approx I$, leading to, $\Lambda^s_{bdacn_1 n_2} \approx s_{\lambda f}(0,t)$. 
With all this, Eq.~\ref{eq:pathintscalarZI6} becomes,
\begin{eqnarray}
	\label{eq:ziz01}
	\frac{Z_I^1}{Z_0} = 	3 \frac{\beta_0^3}{\dV^2}  s_{\lambda f}(0,t)  
	\left ( \left [ \sum_{a} \sum_{n1} \frac{1}{2 D_{n1,a,a}} \right ] \right )^2.
\end{eqnarray}
The value of $D_{n,a,a}$ in Eq.~\ref{eq:Daa}, can be simplified, by recognizing that 
\begin{itemize}
	\item $\frac{\eta_f(0,t)}{\dV} = \langle \eta(\bx,t) \rangle$, 
	\item $1 + \langle \eta(\bx,t) \rangle = \langle s(\bx,t) \rangle $ and,
	\item $1 - \langle \eta(\bx,t) \rangle \approx \frac{1}{\langle s(\bx,t) \rangle}$.
\end{itemize}
Substituting the value of $D_{n,a,a}$ thus obtained, in Eq.~\ref{eq:ziz01},
\begin{multline}
	\label{eq:ziz02}
	\frac{Z_I^1}{Z_0} = 	3 \frac{\beta_0^3}{\dV^2}  s_{\lambda f}(0,t)\\
\times	\left [\sum_a\sum_n \frac{1}{\beta_0^2 \left \{ (\frac{C_n^2}{\langle s(\bx,t) \rangle} + \langle s(\bx,t) \rangle \left (\bp_a^2  +  m^2) \right ) \right \} } 
	\right ]^2.
\end{multline} 
Finally,
\begin{multline}
	\frac{Z_I^1}{Z_0} = 	\frac{3}{4} (\beta_0 \dV)  \langle s(\bx,t) \lambda(\bx,t) \rangle\\ 
\times	\Bigg [ 
	\int \frac{d^3p}{(2\pi)^3} \frac{1}{\omega_p} \coth\left ( \frac{\omega_p \beta_0 \langle s(\bx,t)\rangle}{2} \right ) \Bigg ]^2,
\end{multline} 
where, $\omega_p = \sqrt{\bp^2 + m^2}$ and $\langle s(\bx,t) \lambda(\bx,t) \rangle = s_{\lambda f}(0,t)$.

For, $r=1$, one may then determine, $\langle E(t) \rangle = \frac{\partial \ln(Z(t))}{\partial \beta_0}$.

\section{The Ricci tensor and negative potential}
\label{sec:EFE}
The Einstein field equations are now determined in the 5-D space-time-temperature.
We use the letters, $a$, $b$, as indices for the 5-D space-time, i.e., $a$, $b$ = 0, 1, 2, 3, 4, with the index 0 referring to the temperature dimension, while, the index 1 refers to the time dimension,
We use the letters, $\mu$, $\nu$, as indices for the 4-D Lorentzian space-time, i.e., $\mu$, $\nu$ = 1, 2, 3, 4.
A superscript, $(N)$, within brackets, refers to $N$ dimensional space. For example, $\nabla^{(4)}_{\mu}$, refers to the covariant derivative in 4-D space-time.
Let us consider the Lagrangian,
\begin{equation}
\label{eq:lagscalarR}
\nonumber	\sL(\phi,\partial_a \phi) =  \frac{1}{2} \left ( -\partial_{a}\phi \partial^{a}\phi - m^2\phi^2 - \zeta R \phi^2 \right ),
\end{equation}
with the sign convention (+,-,+,+,+),
and the corresponding action,
\begin{equation}
	S = \int \sqrt{-g^{(5)}} \sL(\phi,\partial_a \phi) d^5x.
\end{equation}
Let us consider the 5-D metric, $g^{(5)}_{ab}$, as:
            \begin{equation}
		    \label{eq:5Dmetric}
		    g^{(5)}_{ab} = 
                \left[ \begin{array}{c c}
			s^2(g^{(4)}_{\mu\nu}(\bx,t),\bx,t)& 0\\
				    0 & g^{(4)}_{\mu\nu}\\
                \end{array} \right ],
            \end{equation}
where, $g^{(4)}_{\mu\nu}$, is the usual metric tensor in 4-D space-time, and is purely due to gravitational fields. 
It is also noted that $s$ would now additionally be a function of $g^{(4)}_{\mu\nu}$ also.
As an example, based on the Ehrenfest Tolman effect~\cite{ET1,ET2}, in the case of a uniform temperature (system in global equilibrium), in a gravitational field, $s(g^{(4)}_{\mu\nu}(\bx,t),\bx,t) = \sqrt{g_{\mu\nu}\zeta^{\mu}\zeta^{\nu}}$, where $\zeta^{\mu}$ is the Killing vector in 4-D space-time. 
We speculate that, the function, $s()$, can be factored, i.e.,
\begin{equation}
	\label{eq:sfactor}
	s(g_{\mu\nu}(\bx,t),\bx,t) = s_1(g_{\mu\nu}(\bx,t))\, s_2(\bx,t),
\end{equation}
where, $s_1(g_{\mu\nu}(\bx,t))$ is due to the Ehrenfest Tolman effect, and $s_2(\bx,t)$ is due to thermal gradients. It is the simplest function that has the following limits:
\begin{enumerate}
	\item When the thermal system is in global equilibrium,  $s(g_{\mu\nu}(\bx,t),\bx,t) \rightarrow s_1(g_{\mu\nu}(\bx,t))$, and,
	\item when $g_{\mu\nu} \rightarrow \eta_{\mu\nu}$, $s(g_{\mu\nu}(\bx,t),\bx,t) \rightarrow  s_2(\bx,t)$.
\end{enumerate}
It is, however desirable to prove or improvise upon Eq.~\ref{eq:sfactor}, based on first principles. 

For the metric, $g^{(5)}_{ab}$, the Einstein field equations in the 5-D space become: 
\begin{equation}
	\label{eq:EFE5D}
	R^{(5)}_{ab} - \frac{1}{2}g^{(5)}_{ab} R^{(5)} = \frac{8\pi G}{c^4} T^{(5)}_{ab},
\end{equation}
where,
\begin{equation}
	T^{(5)}_{ab} = -\frac{2}{\sqrt{g^{(5)}}} \frac{\delta S}{\delta g^{(5)}_{ab}}.
\end{equation}
The 5-D Ricci tensor, $R^{(5)}_{ab}$, can be expressed in terms of the 4-D covariant derivative operator, $\nabla^{(4)}_{\mu}$, and the 4-D Ricci tensor, $R^{(4)}_{\mu\nu}$, as:
            \begin{equation}
		    R^{(5)}_{ab} = 
                \left[ \begin{array}{c c}
			R_{\beta\beta} & 0\\
				    0 & R^{(4)}_{\mu\nu} - \frac{1}{s}\nabla^{(4)}_{\mu} \nabla^{(4)}_{\nu} s\\
                \end{array} \right ].
            \end{equation}

where, 
	$ R_{\beta\beta} = s\nabla^{(4)\mu} \nabla^{(4)}_{\mu} s $.
	The treatment of the geodesic equation in Ref.~\cite{gans6}, gives an insight into the physical interpretation of how thermal gradients, effects an apparent curvature in the Lorentzian space-time.
	It can be easily seen that the expression for $R^{(5)}_{ab}$ reduces to the 4-D Ricci tensor, for the uniform temperature case, i.e., $\partial_{\mu}s\rightarrow 0$. 
	  Another thing to note is that in the metric $s(\bx,t)^2d\beta^2 + g_{\mu\nu}dx^{\mu}x^{\nu}$, the temperature dimension can be eliminated by taking the limit, $s \rightarrow 0$. Thus, for a vacuum, one may take the limits $\partial_{\mu}s \rightarrow 0$, followed by $s\rightarrow 0$, in which case, the Einstein field equations (Eq.~\ref{eq:EFE5D}) reduce to the usual 4-D space-time equations. 
	Finally the 5-D Ricci scalar, $R^{(5)}$ can be expressed as:
	\begin{equation}
		R^{(5)} = R^{(5)a}_{~~a} = R^{(4)} - \frac{2}{s} \nabla^{(4)\mu}\nabla^{(4)}_{\mu} s,
	\end{equation}
	where, $R^{(4)}$ is the conventional 4-D Ricci scalar, in the Lorentzian space, due to gravitational fields, in the absence of any temperature gradients.  
	When the 5-D Ricci scalar is sufficiently negative, it can lead to spontaneous symmetry breaking, i.e., when, $m^2 + \zeta R^{(5)} < 0$. This translates to:
	\begin{eqnarray}
		\label{eq:sconstraint}
		\nonumber		R ^{(5)}< -\frac{m^2}{\zeta}, \\
		\Rightarrow \frac{2}{s} \nabla^{(4)\mu}\nabla^{(4)}_{\mu} s >  R^{(4)} + \frac{m^2}{\zeta}.
	\end{eqnarray}
	It is to be noted that the quantity, $\frac{2}{s} \nabla^{(4)\mu}\nabla^{(4)}_{\mu} s$, is scale-invariant, i.e., the absolute temperature is inconsequential. Only the variations in temperature matter.
	Since, $s(\bx,t)$ is a scalar, $\frac{2}{s} \nabla^{(4)\mu}\nabla^{(4)}_{\mu} s$ simplifies to $\frac{2}{s} \nabla^{(4)\mu}\partial_{\mu} s$.

	As the coupling parameter, $\zeta$, becomes smaller,  $\frac{m^2}{\zeta}$ becomes larger, and the likely hood of $\frac{2}{s} \nabla^{(4)\mu}\nabla^{(4)}_{\mu} s$ satisfying Eq.~\ref{eq:sconstraint} becomes remote. In the extreme case of $\zeta \rightarrow 0$,  Eq.~\ref{eq:sconstraint} can never be satisfied, and spontaneous symmetry breaking can never occur. 
Thus, in the case of a scalar particle, minimally coupled to gravity, i.e., $\zeta \rightarrow 0$, spontaneous symmetry breaking due to thermal gradients can never occur. This is consistent with the fact that the term, $\zeta R \phi^2$, is absent from the Lagrangian in Eq.~\ref{eq:lagscalarR}, and thus precludes the possibility of spontaneous symmetry breaking.
	
	In the case of a thermal medium formed in a collider, e.g., like the QGP at the LHC, the gravitational effects can be neglected. This gives $R^{(4)}=0$.
	Equation~\ref{eq:sconstraint}, then simplifies to:
	\begin{eqnarray}
		\label{eq:sconstraint_collider}
		\frac{2}{s} \partial^{\mu}\partial_{\mu} s >  \frac{m^2}{\zeta}.
	\end{eqnarray}
	One may, however, consider Eq.~\ref{eq:sconstraint_collider}, with caution. For Eq.~\ref{eq:sconstraint_collider} to be satisfied, $\frac{2}{s} \partial^{\mu}\partial_{\mu} s$ needs to be significantly large. But if thermal gradients are very large, the thermal medium may cease to be in local thermal equilibrium, and the current formalism ceases to be valid. 

	The scenario in Eq.~\ref{eq:sconstraint} is however different. To see why, the value of $\frac{2}{s} \nabla^{(4)\mu}\nabla^{(4)}_{\mu} s$ is affected by three factors:
\begin{enumerate}
	\item The thermal gradients, $\partial_{\mu}s_2(\bf{x},t)$ (refer Eq.~\ref{eq:sfactor} for the factorization of $s = s_1 s_2$),
	\item The Ehrenfest Tolman effect, represented by $s_1(\bf{x},t)$, and,
	\item The terms containing the Christoffel symbols, arising as part of the covariant derivative.
\end{enumerate}
Whereas the first factor may be small for a medium in local thermal equilibrium, the second and third factors depend on the strength of the gravitational field. For a thermal system under global equilibrium, $\partial_{\mu} s_1$ can be large if gravitational fields are very strong. The assumption here is that the Ehrenfest Tolman relation holds even when the gravitational fields are strong. 
The term involving the Christoffel symbol can also be large in a strong gravitational field.
Thus, it seems possible, that a thermal system in an exceptionally strong gravitational field, may satisfy Eq.~\ref{eq:sconstraint} and induce spontaneous symmetry breaking.
\section{Summary}
\label{sec:summary}

In this work, it is seen that the concept of 5-D space-time-temperature, permits the incorporation of time-dependent temperature effects, 
for systems that continue to be in local thermal equilibrium.
The bulk thermodynamic property, namely, the energy density expectation value, has been calculated perturbatively, for a thermal bath with both spatial and temporal variations, in the absence of gravity.
The Einstein field equations have been determined in 5-D space-time-temperature, which incorporates both gravitational and thermal effects. 
In the presence of a thermal variation, both temperature gradients and gravity determine the net scalar curvature. 
The net curvature in the 5-D space can be negative. If the d'Alembertian of the temperature variation is sufficiently high, spontaneous symmetry breaking may occur in a scalar field, particularly in the presence of a strong gravitational field.

As mentioned in Ref.~\cite{gans6}, the curvature of the 5-D space due to temperature variations can be validated experimentally, in terrestrial experiments, due to the much lower energy scales involved. Thus, if the concept of the curved 5-D space is confirmed, then it may have important consequences, such as the spontaneous symmetry breaking as described in this work. 
This current formalism may be useful in situations where both gravity and thermal variations are present, like in the interior of a neutron star~\cite{neutron1, neutron2}.
The rotation curves of galaxies have been a subject of intensive research~\cite{galaxy1, galaxy2}.
It might even be possible to consider a galaxy as a thermal medium, with the stars as point particles. However, a galaxy is not in equilibrium, which might pose challenges for such an approach.

\appendix
\section{The 8-D delta function}
\label{sec:8dimsec}
The delta function $\delta(E^2 - \omega^2 - \bf{p}^2 - {\it m}^2)$ is partially Lorentz invariant, and not fully Lorentz invariant.
The quantity, $(E^2 - \bf{p}^2 - {\it m}^2)$, which characterizes a particle, is Lorentz invariant. However, the quantity $\omega^2$ is not. The Matsubara frequency, $\omega$, having the dimensions of energy, will undergo a Lorentz transformation, like energy.
In Ref.~\cite{gans6}, a Lorentz invariant 8-D space $\beta^{\mu}\times x^{\nu}$ was proposed, with $\beta^{\mu} = \beta u^{\mu}$, where, $u^{\mu}$ is the 4-velocity of the thermal bath. The quantity, $x^{\nu}$, is the normal 4-D space-time.
The 8-D space under a Lorentz boost would transform as Ref.~\cite{gans6}:
                \begin{equation}
                \label{eq:8D}
                \left[ \begin{array}{c}
                        \beta'^{\gamma}\\
                        x'^{\delta}\\
                \end{array} \right ]
                         =
                \left[ \begin{array}{c c}
                        \Lambda^{\gamma}_{\alpha}&0\\
                        0&\Lambda^{\delta}_{\beta}\\
                \end{array} \right ]
                \left[ \begin{array}{c}
                        \beta^{\alpha}\\
                        x^{\beta}\\
                \end{array} \right ],
                \end{equation}
where, $\Lambda^{\gamma}_{\alpha}$ and $\Lambda^{\delta}_{\beta}$, represent the Lorentz transformation in a 4-D sub-manifold.
Thus, this 8-D space, which describes both the particle space and the thermal space is fully Lorentz invariant.
Let us construct a 4-Matsubara frequency, $\omega^{\mu} = \omega u^{\mu}$.
The inner products, $\beta^{\mu}\beta_{\mu}$, $\beta^{\mu} \omega_{\mu}$ and $\omega^{\mu}\omega_{\mu}$ are easily seen to be Lorentz invariant.
In this proposed 8-D space, the constraining delta function would be: 
\begin{equation}
\label{eq:delta8D}
	\delta(E^2 - \omega_0^2 - \bf{p}^2 + {\bm \omega}^2 - {\it m}^2),
\end{equation}
where, $\omega_{\mu}$ is split as $(\omega_0, {\bf \omega })$.
To understand the physical intuition behind the 8-D delta function, let us refer to Fig.~\ref{fig:8dimfig}. Figure~\ref{fig:8dimfig} depicts a particle with 4-momentum $(E,\bf{p})$,  and mass $m$, traversing a thermal medium, with inverse temperature $\beta$ and 3-velocity $\bf{u}$.
Let the 3-velocity of the particle corresponding to the 3-momentum, $\bf{p}$, be $\bf{v}$.
The particle interacts with the thermal medium and undergoes a decay (or enhancement). As it decays, the decayed fraction of the particle merges with the thermal medium, and as it merges, it acquires a velocity expectation value of $\bf{u}$, which is identical to the thermal medium. 
This leads to an alteration of the momentum of the particle, and this alteration of the momentum is dependent on the degree of decay (enhancement) of the particle. 
Consequently, what needs to appear in a conservation equation is not just the 3-momentum, $\bp$, but a combination of the momentum of the particle and its decayed (enhancement) fraction.
The degree of decay (enhancement), in turn, depends on the degree of interaction of the thermal medium with the particle, which is characterized by the Matsubara frequency, $\omega$.
It is seen that the quantity, $\omega_{\mu}$, is proportional to both $\omega$ and the 4-velocity, $u_{\mu}$.
The delta function in Eq.~\ref{eq:delta8D} encapsulates the above scenario, by incorporating $\omega_{\mu}$.
It can be noted that when $\omega=0$, all interaction with the thermal medium disappears. In fact, $\omega=0$, corresponds to $\beta \rightarrow \infty$, i.e., zero temperature or vacuum. 

Let us recap the argument in Sec.~\ref{sec:timevariation}. The intrinsic energy of a particle is $E$, and $\omega = iE_c$, the interaction energy of the particle with the thermal medium.
The magnitude of total energy is then $= \sqrt{E^2 + E_c^2} = \sqrt{E^2 - \omega_0^2}$. On similar lines, let us call $\bf{q} = -{\it i}\bm{\omega}$ (=$-i\omega \bf{u}$), as the momentum of the decayed part of the particle. The net momentum = $ \sqrt{\bf{p}^2 + \bf{q}^2} = \sqrt{\bf{p}^2 - \bm{\omega}^2}$. The mass shell equation then becomes:
\begin{equation}
	\label{eq:massshell8d}
	E^2 - \omega_0^2 = (\bf{p}^2 - \bm{\omega}^2) + {\it m}^2.
\end{equation}
Alternatively,
\begin{equation}
	\label{eq:conservation8D}
	E^2 - \omega_0^2 - \bf{p}^2 + \bm{\omega}^2 - {\it m}^2 = 0.
\end{equation}
Equation ~\ref{eq:conservation8D} is fully Lorentz invariant and can represent the momentum conservation of a relativistic particle in a relativistic thermal medium.
The 5-D delta function, $\delta(E^2 - \omega_0^2 - {\bf p}^2  - m^2)$, is a special case of Eq.~\ref{eq:conservation8D}, when $\bf{u=0}$.
If, however, one were to consider cases such as the cosmic microwave background (CMB) as a thermal bath, $\bf{u}$ may be irrelevant, and the 5-D delta function might be sufficient.


\begin{thebibliography}{99}

\bibitem{matsubara} Matsubara, T.  A New Approach to Quantum-Statistical Mechanics. Progress of Theoretical Physics, {\bf 14}, 4, 351–378 (1955).
\bibitem{martin} Martin, P. C., Schwinger, J. 
Physical Review, 115(6), 1342–1373 (1959). 
\bibitem{gorkov} A.A. Abrikosov, L.P. Gor'kov, I.E. Dzyaloshinskii
JETP, Vol. 9, No. 3, p. 636 (1959)
\bibitem{adas} Arnold, P., Vokos, S., Bedaque, P., Das, A. 
	Physical Review D, 47(10), 4698 (1993).
\bibitem{misc1} Y. Aoki, G. Endrodi, Z. Fodor, S. D. Katz, and K. K. Szabo, 
Nature 443, 675-678 (2006), arXiv:hep-lat/0611014.
\bibitem{misc2} Kenji Fukushima and Tetsuo Hatsuda, 
Rept. Prog. Phys.  74, 014001 (2011), arXiv:1005.4814.
\bibitem{Kraemmer} Kraemmer and Rebhan, Rept.Prog.Phys. {\bf 67} (2004).
\bibitem{Kraemmer2} Kraemmer, Rebhan and H. Hchulz, Annals. of Phys. {\bf 238} (1995).
\bibitem{misc3} Heng-Tong Ding, Frithjof Karsch, and Swagato Mukherjee, 
Int. J. Mod. Phys. E24, 1530007 (2015), arXiv:1504.05274.
\bibitem{misc7} Blaizot J. P. and Iancu E., 
Phys. Rep., {\bf 359} 355–528 (2002); arxiv hep-ph/0101103.
\bibitem{misc4} David E. Morrissey and Michael J. Ramsey-Musolf, 
New J. Phys. 14, 125003 (2012), arXiv:1206.2942.
\bibitem{misc5} Andrew G. Cohen, D. B. Kaplan, and A. E. Nelson, 
Ann. Rev. Nucl. Part. Sci. 43, 27-70 (1993), arXiv:hep-ph/9302210.
\bibitem{misc6} V. A. Rubakov and M. E. Shaposhnikov, 
Usp. Fiz. Nauk 166, 493-537 (1996), arXiv:hep-ph/9603208.
\bibitem{tft} Yasushi Takahashi, Hiroomi Umezawa, International journal of Modern Physics B, {\bf 10} (1996).
\bibitem{epja} Torbjorn Lundberg and Roman Pasechnik, European Physical Journal A,  {\bf 57}, (2021).
\bibitem{nature}  P. Braun-Munzinger and Johanna Stachel, Nature {\bf 448}, 302-309 (2007). 
\bibitem{naturephy}  PHENIX Collaboration, Nature Physics {\bf 12}, (2018); arxiv:nucl-ex/1805.02973 (2018).
\bibitem{gans5}  S. Ganesh and M. Mishra,  Progress of Theoretical and Experimental Physics, {\bf 2021}, 1, 013B09 (2021).
\bibitem{gans6}  S. Ganesh ,  Int. J. Mod. Phys. A, 
	{\bf 37}, 17, 2250125 (2022);
	arXiv:hep-th/2206.13324 (2022).
\bibitem{ssb1} S. Coleman, E. Weinberg, Phys. Rev. D, {\bf 7}, 4 (1973).
\bibitem{ssb2} J. Baglio, A. Djouadi, JHEP, {\bf 2011}, 55 (2011); arXiv: hep-ph/1012.0530 (2012).
\bibitem{higgs1} Peter Higgs, 
	Physics Letters 12, 132 (1964).
\bibitem{higgs2} Peter W. Higgs, 
	Phys. Rev. Lett. 13, 508 (1964).
\bibitem{goldstone} J. Goldstone, A Salam and S Weinberg, 
	Physical Review, 127, 965 (1962).
\bibitem{kapusta} J. Kapusta, C. Gale, "Finite Temperature Field Theory and Applications", Cambridge University Press, 2e (2006).
	\bibitem{Nbeta} C. W. Bernard, {\it Phys. Rev. D}, {\bf 9}(12), 3312 (1974).
\bibitem{eigen} Yuji Nakatsukasa,
Lecture Notes in Computer Science, EPASA 2015, pp 233-249.
\bibitem{ET1} R. C. Tolman,
Phys. Rev. {\bf 35} 904–924 (1930).
\bibitem{ET2}  R. C. Tolman and P. Ehrenfest,
Phys. Rev.  {\bf 36}, 12, 1791–1798 (1930).
\bibitem{neutron1} N Andersson, GL Comer, and K Glampedakis. 
	Nuclear Physics A, 763, 212–229 (2005), 
\bibitem{neutron2}Petarpa Boonserm, Matt Visser, and Silke Weinfurtner. 
	Phys. Rev. D, {\bf 76}, 4, 044024 (2007), 
\bibitem{galaxy1} Y. Sofue et. al., "The Astrophysical Journal", {\bf 523}, 136-146 (1999). 
\bibitem{galaxy2} Kyu-Hyun Chae et. al., 
	The Astrophysical Journal, {\bf 904} (2020).

\end{thebibliography}
\end{document}